\documentclass[intlimits,sumlimits,12pt]{iopart}

\expandafter\let\csname equation*\endcsname\relax
\expandafter\let\csname endequation*\endcsname\relax

\newcommand{\inte}{\int\limits}

\newcommand{\produ}{\prod\limits}

\newcommand{\lime}{\lim\limits}
\newcommand{\diag}{\textrm{diag\,}}

\def\newblock{\hskip .11em plus.33em minus.07em}

\usepackage{iopams}
\usepackage{amsmath}
\usepackage{dsfont}
\usepackage{cite}
\usepackage{graphicx}
\usepackage[numbers]{natbib}

\begin{document}

\title[Matrix Moments]{Matrix Moments in a Real, Doubly Correlated Algebraic Generalization
  of the Wishart Model}

\author{Thomas Guhr and Andreas Schell}

\address{Fakult\"at f\"ur Physik, Universit\"at Duisburg--Essen, Duisburg, Germany}
\ead{thomas.guhr@uni-due.de}
\vspace{10pt}

\begin{abstract}
  The Wishart model of random covariance or correlation matrices
  continues to find ever more applications as the wealth of data on
  complex systems of all types grows. The heavy tails often
  encountered prompt generalizations of the Wishart model, involving
  algebraic distributions instead of a Gaussian. The mathematical
  properties pose new challenges, particularly for the doubly
  correlated versions. Here we investigate such a doubly correlated
  algebraic model for real covariance or correlation matrices. We
  focus on the matrix moments and explicitly calculate the first and
  the second one, the computation of the latter is non--trivial.  We
  solve the problem by relating it to the Aomoto integral and by
  extending the recursive technique to calculate Ingham--Siegel
  integrals. We compare our results with the Gaussian case.
\end{abstract}

%
\vspace{2pc}
\noindent{\it Keywords}: random matrix theory, statistical inference, algebraic heavy tails

\submitto{\JPA}

%
%

\section{Introduction}
\label{sec0}

The Wishart model for random correlation or covariance matrices plays
a central r\^ole in statistical inference.  Its original
version~\cite{Wishart1928}, correlates the rows of the random data
matrices, \textit{i.e.}, the time series, more
recently~\cite{Simon2004,Burda2005,McKay2007,Waltner2014,Burda2015}
correlations of the columns, \textit{i.e.}, of the position series,
were also included. A crucial assumption is that a Gaussian form of
this multi--multivariate distribution is justified, a viewpoint that
can often be backed by arguments related to the Central Limit
Theorem. Nevertheless, the enormous amount of data on various systems
revealed that such a line of reasoning has its limitations.  Finance
is probably the field with the best evidence, often revealed by
physicists who also produced a large body of work on random matrices
in this context, including non--Gaussian
models~\cite{bouchaud2000theory,Laloux1999,Laloux2000,Plerou1999a,Plerou2002,Pafka2004,potters2005financial,drozdz2008empirics,Kwapien2006,biroli2007student,Burda2001,Burda2002,Akemann2008,burda2011applying}. It
remains a challenge to systematically explore statistical properties
of other systems in that respect and to that extent.

Here, we study an algebraic extension of the doubly correlated Wishart
model that we used recently~\cite{GS2020a}. It involves a determinant
and thus all matrix invariants, while the Gaussian Wishart model
relies on the trace only. Our model~\cite{GS2020a} generalizes
previously introduced ones by Forrester and Krishnapur~\cite{FK2009}
as well as by Wirtz, Waltner, Kieburg and Kumar~\cite{WWKK2016}.
Importantly, we consider real matrices, \textit{i.e.} the case also
referred to as orthogonal, because it is the most commonly encountered
one in the application to data. We mention in passing that
Ref.~\cite{GS2020a} also extends a new
interpretation~\cite{Schmitt2013,Meudt2015} of the Wishart model,
namely a random matrix model for non--stationarity, conceptually
different from the usage in statistical inference.  Various
multivariate algebraic amplitude distributions were modeled and
calculated.

Not surprisingly, an algebraic extension of the doubly correlated
Wishart model is mathematically considerably more complicated than the
Gaussian version, triggering the present study. The difficulties are
particularly severe, as we employ the orthogonal, not the much more
convenient unitary symmetry. We compute the first and the second
matrix moment of our algebraic model, as they can be used in data
anlysis to fix parameters of the above mentioned multivariate
algebraic amplitude distributions and thus also of the algebraic
distribution for the random correlation or covariance matrices.  The
second moment poses serious challenges which we face by further
developing some techniques, more precisely, we manage to establish
helpful relations to the Aomoto integral~\cite{Aomoto_1987,Mehta_1997}
and, furthermore, we extend the recursive
method~\cite{Siegel_1935,Fyodorov_2002} to compute integrals related
to the Ingham--Siegel type. The strategy is to avoid or circumvent
group integrals, which are, as is well known, quite cumbersome for the
orthogonal symmetry relevant here, much more complicated as for the
unitary one. We have two goals. First, we wish to present explicit
formulae for the first two moments and the resulting matrix variance
to be used in data anlysis. We also compare with the Gaussian
case. Second, we wish to contribute new pieces to the tool box of
Random Matrix Theory, which are useful for algebraic extensions of the
Wishart model and in other random matrix models.

The paper is organized as follows. In Sec.~\ref{sec1}, we define the
problem mathematically and set up proper generating functions. In
Sec.~\ref{sec2}, we compute the first matrix moment, preparing for the
calculation of the much more involved calculation for the second
matrix moment in Sec.~\ref{sec3}. We conclude in Sec.~\ref{sec4}.
Some details are worked out in the Appendix.

\section{Determinantal Distribution, Matrix Moments and Generating Function}
\label{sec1}

In Sec.~\ref{sec11}, we introduce our algebraic extension of the
doubly correlated Wishart model. In Sec.~\ref{sec12}, we define the
matrix moments and introduce as well as compute a generating function.

\subsection{Determinantal Distribution}
\label{sec11}

Consider $K \times N$ real data matrices $X$ with elements $X_k(n)$ If
one views the $K$ rows of $X$ as time series and its $N$ columns as
position series, the $K \times K$ and $N \times N$ matrices
\begin{align}
 \frac{1}{N} XX^\dagger \qquad \text{and} \qquad \frac{1}{K} X^\dagger X 
  \label{sec1.1}
\end{align}
are the sample covariance matrices of time and position series,
respectively. We always uese the dagger symbol to indicate the
transpose.  In a statistical model, one draws the data matrices $X$
from a distribution. There are many such random matrix applications in
the theory of complex systems. In recent years one became interested
in algebraic distributions of covariances or correlations to model
heavy tails found in data.  A nearlying choice is the determinantal
distribution
\begin{align}
  w_A (X|\Sigma,\Xi) &= \frac{\alpha_{KNLM}}{\displaystyle{\det}^L \left( \mathds{1}_N +\frac{1}{M}\Xi^{-1} X^\dagger \Sigma^{-1}X \right)} \nonumber\\
  &= \frac{\alpha_{KNLM}}{\displaystyle{\det}^L \left( \mathds{1}_N +\frac{1}{M}\Xi^{-1/2} X^\dagger \Sigma^{-1}X\Xi^{-1/2} \right)}
  \label{eq:DetVerX}\\
  \alpha_{KNLM} &= \frac{1}{\sqrt{\det 2\pi \Xi\otimes\Sigma}}\sqrt{\frac{2}{M}}^{KN}
  \prod_{n=1}^{N} \frac{\Gamma(L-(n-1)/2))}{\Gamma(L-(K+n-1)/2)} \ .
  \label{eq:DetVerXnorm}
\end{align}
This distribution exists, in the sense that it is integrable, if the
condition
\begin{align}
	L& > \frac{N+K-1}{2} 
\end{align}
holds. The distribution~\eqref{eq:DetVerX} was introduced in
Ref.~\cite{GS2020a}, generalizing related ones defined in
Refs.~\cite{FK2009,WWKK2016}.  We give the
distribution~\eqref{eq:DetVerX} in two equivalent forms in
Eq.~\eqref{eq:DetVerX}, one with the argument of the determinant being
real--symmetric, and another one, in which this is not the case. The
fixed matrices $\Sigma$ and $\Xi$ are positive definite of dimensions
$K \times K$ and $N \times N$. They are, as to be discussed below,
different from, but related to the corresponding covariance matrices.
The distribution~\eqref{eq:DetVerX} depends on two shape parameters
$L$ and $M$.  It converges to the Gaussian
\begin{align}
  w_G (X|C,D) &= \frac{1}{\sqrt{\det 2\pi \Xi\otimes\Sigma}}\exp\left(-\frac{1}{2}\tr \Xi^{-1} X^\dagger \Sigma^{-1}X \right) \ ,
  \label{eq:GauVerX}
\end{align}
known as the doubly correlated Wishart
distribution~\cite{Wishart1928,Simon2004,Burda2005,McKay2007,Waltner2014,Burda2015},
if $L$ and $M$ are taken to infinity under the condition
\begin{align}
\lim_{L,M\to\infty} \frac{M}{L} &= 2 \ .
\label{eq:GauVerR2}
\end{align}
It is seldom possible to directly compare a distribution such
as~\eqref{eq:DetVerX} or~\eqref{eq:GauVerX} with data, one usually
analyzes other observables $f(\rho|X)$ which depend on some arguments
$\rho$ and parametrically on the time or position series. To carry out
the data comparison, one calculates the ensemble average
\begin{align}
\langle f\rangle_Y(\rho|\Sigma,\Xi) &= \int f(\rho|X) w_Y(X|\Sigma,\Xi) \textrm{d}[X] \ , \qquad Y=G,A \ ,
\label{rmteac}
\end{align}
in which the invariant measure or volume element reads
\begin{align}
\textrm{d}[X] &= \prod_{k=1}^K \prod_{n=1}^N \textrm{d}X_k(n) \ .
\label{rmtvol}
\end{align}
Obviously, the parameters $L$ and $M$ in the algebraic case must have
values that ensure the convergence of the integral~\eqref{rmteac}.

\subsection{Matrix Moments and Generating Function}
 \label{sec12}

To make this program meaningful, one has to wisely choose the matrices
$\Sigma$ and $\Xi$ or, even better, to determine them from data. If
the distribution~\eqref{eq:DetVerX} were Gaussian, this would be an
easy task, because $\Sigma$ and $\Xi$ could simply be obtained from
the data as the sample covariance matrices. This is not so in the
present algebraic case. To determine the matrices $\Sigma$ and $\Xi$,
we have to find out their relation to the covariance matrices. This
motivates the study of the $\nu$--th matrix moments
\begin{align}
  \left\langle \left(\frac{1}{N} XX^\dagger\right)^\nu\right\rangle_Y &= \int \left(\frac{1}{N} XX^\dagger\right)^\nu w_Y(X|\Sigma,\Xi) \textrm{d}[X] \ ,
\label{mom1t}\\
  \left\langle \left(\frac{1}{K} X^\dagger X\right)^\nu\right\rangle_Y &= \int \left(\frac{1}{K} X^\dagger X\right)^\nu w_Y(X|\Sigma,\Xi) \textrm{d}[X] \ , 
\label{mom1p}
\end{align}
for $Y=G,A$, and provided they exist in the algebaric case. Due to the
algebraic structure of $w_A(X|\Sigma,\Xi)$, there are highest possible
matrix moments, depending on the parameters of the distribution.  The
first matrix moments, $\nu=1$, yields the desired relations to the
covariance matrices for the time and position series, respectively,
where the left hand sides are to be identified with the sample
covariance matrices.

The evaluation of the matrix integrals~\eqref{mom1t} and~\eqref{mom1p}
is possible for $\nu=1$ by a proper change of the integration matrix
which makes the $\Sigma$ and $\Xi$ dependence trivial and thus the
remaining integrals much easier.  The calculation of the higher order
moments, $\nu=2,3,\ldots$, turns out to be an interesting and
non--trivial challenge in mathematical physics. Here, we tackle the
calculation of the second matrix moment. For $\nu > 1$, the above
mentioned change of the integration matrix does not remove $\Sigma$ or
$\Xi$ from the relevant integrals. An alternative approach is called
for.  We introduce the generating function
\begin{align}
  Z_Y(J) &= \int \exp\left(-\tr \frac{1}{N}XX^\dagger J\right) w_Y(X|\Sigma,\Xi) \text{d}[X] \ , \qquad Y=G,A \ ,
  \label{eq:ErzFun}
\end{align}
depending on the $K\times K$ real--symmetric source matrix $J$. This
generating function is normalized,
\begin{align}
  Z_Y(0) &= 1 \ .
  \label{eq:ErzFunN}
\end{align}
We define the matrix gradient corresponding to the source matrix,
\begin{align}
	\frac{\partial}{\partial J} &= \begin{bmatrix}
\frac{\partial}{\partial J_{11}} & \frac{1}{2}\frac{\partial}{\partial J_{12}} & \cdots & \frac{1}{2}\frac{\partial}{\partial J_{1K}} \\
\frac{1}{2}\frac{\partial}{\partial J_{12}} & \frac{\partial}{\partial J_{22}} & \cdots & \frac{1}{2} \frac{\partial}{\partial J_{2K}} \\
\vdots & \vdots & \ddots & \vdots \\
\frac{1}{2}\frac{\partial}{\partial J_{1K}}& \frac{1}{2}\frac{\partial}{\partial J_{2K}}  & \cdots & \frac{\partial}{\partial J_{KK}} \\
        \end{bmatrix} \ ,
        \label{eq:AblOpeDef}
\end{align}
in which the factors $1/2$ account for the symmetry of $J$ and $XX^\dagger/N$. We have
\begin{align}
 \left\langle \left(\frac{1}{N} XX^\dagger\right)^\nu\right\rangle_Y
  &= (-1)^\nu \frac{\partial^\nu Z_Y(J)}{\partial J^\nu} \bigg\vert_{J=0} \ , \qquad \nu=0,1,2,\ldots \ .
  \label{eq:AblOpe}
\end{align}
The function $Z_Y(J)$ generates the matrix moments~\eqref{mom1t} and is
defined similar to a characteristic function. We notice that the
normalization~\eqref{eq:ErzFunN} is scalar and thus not identical to
the zeroth moment which is equal to the unit matrix $\mathds{1}_K$. To
generate the matrix moments~\eqref{mom1p}, $XX^\dagger/N$ has to be
replaced by $X^\dagger X/K$ and $J$ has to be a real--symmetric
$N\times N$ matrix. However, once the moments~\eqref{mom1t} are
computed, the moments~\eqref{mom1p} can be inferred without further
calculation by replacing parameters and dimensions.

A natural object involving the first and the second matrix moments is
the matrix variance
\begin{align}
  {\textrm{var}_Y}\left(\frac{1}{N} XX^\dagger\right)
  &= \left\langle\left(\frac{1}{N} XX^\dagger-\left\langle\frac{1}{N} XX^\dagger\right\rangle_Y\right)^2\right\rangle_Y \nonumber\\
  &= \left\langle\left(\frac{1}{N} XX^\dagger\right)^2\right\rangle_Y - \left\langle\frac{1}{N} XX^\dagger\right\rangle_Y^2 \ ,
  \label{sec2.50}
\end{align}
as it is, among other things, invariant under a shift by a fixed matrix $\Omega$, say,
\begin{align}
  {\textrm{var}_Y}\left(\frac{1}{N} XX^\dagger + \Omega\right)
  &=   {\textrm{var}_Y}\left(\frac{1}{N} XX^\dagger\right) \ .
  \label{sec2.51}
\end{align}
To avoid confusions, we emphasize that Eq.~\eqref{sec2.50} defines the
variance of the matrix $XX^\dagger/N$ that serves as estimator for the
sample covariances.  Hence, it defines the variance of the
covariances, and its dimension is that of the time series to the fourth
power. Nevertheless, it has all the properties expected for a
variance, in particular it is by construction a positive semidefinite
matrix.

We first consider the algebraic case and use the generating function
to cast the matrix integrals~\eqref{mom1t} into a form better adjusted
for explicit evaluation. To this end, we employ the Ingham--Siegel
integral~\cite{Siegel_1935,Fyodorov_2002}
\begin{align}
  \int_{S > 0} \exp(-\tr SR) {\det}^{{q-(N+1)/2}} S \textrm{d}[S]
      &= \frac{\pi^{N(N-1)/4}\prod_{n=1}^{N}\Gamma(q-(n-1)/2)}{\det ^q R} \ ,
\label{eq:IngSieInt}
\end{align}
where the matrices $S$ and $R$ are $N\times N$ real--symmetric, $S$
has to be positive, indicated by $S>0$. Convergence is guaranteed if
$q\ge (N+1)/2$. We write $w_A(X|\Sigma,\Xi)$ as an Ingham--Siegel
integral where we take the real--symmetric matrix argument of the
determinant in the second form of Eq.~\eqref{eq:DetVerX}. The entire
dependence on $X$ and $X^\dagger$ in the exponent can be written as
\begin{align}
  -\tr X^\dagger\frac{J}{N}X - \tr \Xi^{-1/2}\frac{S}{M}\Xi^{-1/2}X^\dagger \Sigma^{-1} X
  &= - x^\dagger\left(\mathds{1}_N\otimes\frac{J}{N}+\Xi^{-1/2}\frac{S}{M}\Xi^{-1/2}\otimes\Sigma^{-1}\right)x \ ,
\label{sec1.2}
\end{align}
owing to the relation $\tr FX^\dagger GX = x^\dagger (F^\dagger
\otimes G) x$ in which $x$ is a $KN$ component vector, constructed
from the columns $X(n), \ n=1,\ldots,N$ of the $K\times N$ matrix $X$.
The real matrices $F$ and $G$ have dimensions $N\times N$ and $K\times
K$, respectively. The integral over $X$ is then of simple Gaussian
form.  Thus we find
\begin{align}
  Z_A(J) &=  \frac{1}{\pi^{N(N-1)/4}M^{KN/2}\sqrt{\det\Xi\otimes\Sigma}\prod_{n=1}^N\Gamma(L-(K+n-1)/2)} \nonumber\\
 & \qquad\qquad  \int_{S > 0} \frac{\exp(-\tr S) {\det}^{{L-(N+1)/2}} S}
             {\sqrt{\det(\Xi^{-1/2}S\Xi^{-1/2}/M\otimes\Sigma^{-1}+\mathds{1}_N\otimes J/N)}}
                                                      \textrm{d}[S] \ .
\label{sec1.3}
\end{align}
Expanding the large determinant expression in the integrand to second order in $J$,
\begin{align}
& \frac{1}{\sqrt{\det(\Xi^{-1/2}S\Xi^{-1/2}/M\otimes\Sigma^{-1})}}
  \exp\left(-\frac{1}{2}\tr\ln\left(\mathds{1}_N\otimes\mathds{1}_K
                            +M \Xi^{1/2}S^{-1}\Xi^{1/2}\otimes\Sigma J/N\right)\right) \nonumber\\
  & \qquad = \frac{M^{KN/2}}{{\det}^{{K/2}}(\Xi^{-1/2}S\Xi^{-1/2}){\det}^{{N/2}}\Sigma} \nonumber\\
  & \qquad\qquad\qquad \exp\left(-\frac{M}{2N}\tr\Xi S^{-1}\tr\Sigma J
                          + \frac{M^2}{4N^2}\tr(\Xi S^{-1})^2\tr(\Sigma J)^2+{\cal O}(J^3)\right) \nonumber\\
  & \qquad = \frac{M^{KN/2}\sqrt{\det\Xi\otimes\Sigma}}{{\det}^{{K/2}} S}
                                   \left(1-\frac{M}{2N}\tr\Xi S^{-1}\tr\Sigma J \right. \nonumber\\
  & \qquad\qquad\qquad \left. + \frac{M^2}{8N^2}{\tr}^{2}(\Xi S^{-1}){\tr}^2(\Sigma J)+
                         \frac{M^2}{4N^2}\tr(\Xi S^{-1})^2\tr(\Sigma J)^2+{\cal O}(J^3)\right) \ ,
\label{sec1.4}
\end{align}
we arrive at
\begin{align}
  Z_A(J) &=  1 - \frac{1}{\pi^{N(N-1)/4}\prod_{n=1}^N\Gamma(L-(K+n-1)/2)}  \nonumber\\
  & \qquad\qquad  \int_{S > 0} \textrm{d}[S] \exp(-\tr S) {\det}^{{L-(N+K+1)/2}} S
                                           \left( \frac{M}{2N}\tr\Xi S^{-1}\tr\Sigma J \right.  \nonumber\\
  & \qquad\qquad\qquad  - \left. \frac{M^2}{8N^2}{\tr}^{2}(\Xi S^{-1}){\tr}^2(\Sigma J)
                                           - \frac{M^2}{4N^2}\tr(\Xi S^{-1})^2\tr(\Sigma J)^2\right) 
                                                          + {\cal O}(J^3) \ ,
\label{sec1.5}
\end{align}
up to second order in $J$.  We observe that the invariance of the
integration measure allows us to replace the matrix $\Xi$ by the
diagonal matrix
\begin{align}
\Theta &= \diag(\Theta_1,\ldots,\Theta_N)
\label{sec1.a}
\end{align}
of its eigenvalues. This is also true for the generating
function~\eqref{sec1.3} to all orders in $J$, but only form now on, it
will lead to simplifications.

We also compute the Gaussian case for later comparison. It is easily
inferred from the above derivation. We find
\begin{align}
  Z_G(J) &= \frac{1}{\sqrt{\det(\mathds{1}_N\otimes\mathds{1}_K + 2\Xi\otimes \Sigma J/N)}} \nonumber\\
  &= 1 - \frac{1}{N}\tr\Xi \, \tr\Sigma J
          + \frac{1}{2N^2} {\tr}^2\Xi \, {\tr}^2(\Sigma J) + \frac{1}{N^2}\tr\Xi^2\tr(\Sigma J)^2
                            + {\cal O}(J^3) \ ,
\label{sec1.6}
\end{align}
which trivially also only depends on the eigenvalues of $\Xi$.

\section{First Matrix Moment}
\label{sec2}

There are various straightforward ways to calculate the first matrix
moment, including the one mentioned in Sec.~\ref{sec12}. Here, we
choose a method that prepares for the much more demanding computations
of the second matrix moment in Sec.~\ref{sec3}. In Sec.~\ref{sec21} we
reduce the non--invariant matrix integral to be evaluated to an
invariant one, allowing us to apply the Aomoto integral in
Sec.~\ref{sec22} which quickly yields the final result.

\subsection{Reduction to an Invariant Integral}
\label{sec21}

We consider the algabraic case and apply Eq.~\eqref{eq:AblOpe} for
$\nu=1$ to our formula~\eqref{sec1.5}.  The matrix gradient of
$\tr\Sigma J$ is simply the matrix $\Sigma$, and we find for the first
matrix moment
\begin{align}
  \left\langle\frac{1}{N} XX^\dagger\right\rangle_A &= 
                \frac{M\Phi_1(\Xi)}{2N\pi^{N(N-1)/4}\prod_{n=1}^N\Gamma(L-(K+n-1)/2)}\Sigma
\label{sec2.1}
\end{align}
with the function 
\begin{align}
 \Phi_1(\Xi) &= \int_{S > 0} \textrm{d}[S] \exp(-\tr S) {\det}^{{L-(N+K+1)/2}} S \tr\Theta S^{-1} \ .
\label{sec2.2}
\end{align}
The advantage of working with the generating function instead of
evaluating Eq.~\eqref{mom1t} starts becoming visible. The matrix
structure of the first moment results directly from the matrix
gradient, the remaining integral is scalar. While this simplification
is not decisive for the first matrix moment, it will turn out very
helpful for the second one.  We write the trace $\tr\Theta S^{-1}$ as
sum, pull out the $\Theta_n$ from the integral and are left with $N$
integrals, each containing one diagonal element
$[S^{-1}]_{nn}$. Obviously, they cannot depend on the indices $n$,
because with simple changes of variables by permuting the basis, we
can map any diagonal element of $S^{-1}$ on any other one. Put
differently, all integrals must give the same result. Hence we may
make the follwing replacement under the integral
\begin{align}
  \tr\Theta S^{-1} &= \sum_{n=1}^N \Theta_n [S^{-1}]_{nn} \longrightarrow
       \sum_{n=1}^N \Theta_n  \frac{1}{N}\sum_{m=1}^N [S^{-1}]_{mm} = \frac{\tr\Theta}{N} \tr S^{-1} 
\label{sec2.4}
\end{align}
and find
\begin{align}
  \Phi_1(\Xi) &= \frac{\tr\Theta}{N} \int_{S > 0} \textrm{d}[S] \exp(-\tr S) {\det}^{{L-(N+K+1)/2}} S \tr S^{-1} 
\label{sec2.5}
\end{align}
with $\tr\Theta=\tr\Xi$. Importantly, the integral in
Eq.~\eqref{sec2.5} is invariant and may be written as an integral over
the eigenvalues of $S$ only, the integral over the orthogonal group
which diagonalizes $S$ is trivial. In contrast, the original integral
in Eq.~\eqref{sec2.2} also requires a non--trivial integration over
the group and is thus much more complicated. In other words, we
achieved a decoupling of the non--Markovian effects. We change to
eigenvalue--angle coordinates
\begin{align}
S &= UsU^\dagger \qquad \textrm{with} \qquad s = \diag(s_1,\ldots,s_N) \ ,
\label{sec2.6}
\end{align}
where $s_n>0$, the matrix $U$ parameterizes the orthogonal group. The
volume element reads
\begin{align}
  d[S] = \frac{\pi^{N(N+1)/4}}{N!\prod_{n=1}^N\Gamma(n/2)}|\Delta_N(s)|d[s]d\mu(U)
         \qquad \textrm{with} \qquad \Delta_N(s) = \prod_{n<m}(s_n-s_m)
\label{sec2.7}
\end{align}
being the Vandermonde determinant. The invariant Haar measure
$d\mu(U)$ is normalized to unity. We arrive at
\begin{align}
  \Phi_1(\Xi) &= \frac{\pi^{N(N+1)/4}}{N!\prod_{n=1}^N\Gamma(n/2)}
                  \frac{\tr\Xi}{N} \int_{s > 0} \textrm{d}[s] |\Delta_N(s)| \exp(-\tr s) {\det}^{{L-(N+K+1)/2}} s \tr s^{-1} \ ,
\label{sec2.8}
\end{align}
which reduces the problem to the calculation of an $N$ dimensional
eigenvalue integral.

\subsection{Application of the Aomoto Integral and Final Result}
\label{sec22}

The integral in Eq.~\eqref{sec2.8} can be worked out with various
techniques, we find it convenient to apply the Aomoto
integral~\cite{Aomoto_1987,Mehta_1997}. The calculation is carried out
in~\ref{app1} and we find
\begin{align}
  \Phi_1(\Xi) &= \frac{2\pi^{N(N-1)/4}\tr\Xi}{2L-1-(K+N)}\prod_{n=1}^N\Gamma(L-(K+n-1)/2)
\label{sec2.9}
\end{align}
which eventually yields
\begin{align}
  \left\langle\frac{1}{N} XX^\dagger\right\rangle_A &= 
                \frac{M}{2L-1-(K+N)} \frac{\tr\Xi}{N} \, \Sigma
\label{sec2.10}
\end{align}
for the first matrix moment. The singularity in the prefactor is a
clear indication that the first moment only exists if the condition
\begin{align}
  L & > \frac{K+N+1}{2}
\label{sec2.11}  
\end{align}
is fullfilled. With the choice $M=2L-1-(K+N)$ we can enforce existence
of the first matrix moment which then coincides exactly with the result
\begin{align}
  \left\langle\frac{1}{N} XX^\dagger\right\rangle_G &= \frac{\tr\Xi}{N} \, \Sigma
\label{sec2.12}
\end{align}
for the Gaussian model for random covariances or correlation
matrices. Hence, the setting $M=2L-1-(K+N)$ allows us to interpret
$\Sigma$ directly as the (average) sample covariance matrix when
comparing with data.

\section{Second Matrix Moment}
\label{sec3}

In Sec.~\ref{sec31}, we decouple the non--Markovian effects,
\textit{i.e.}, one of the fixed input matrices, from the matrix
integrals. In Sec.~\ref{sec32}, we trace two of the three resulting
integrals back to invariant ones and infer them from the Aomoto
integral. As the problem cannot be fully solved in this way, we carry
out a non--trivial extension of the recursive technique to calculate
integrals of the Ingham--Siegel type in Sec.~\ref{sec33}. We give our
final results, including the matrix variance, and compare with the
Gaussian case in Sec.~\ref{sec34}.

\subsection{Matrix Integral and Decoupling of the Non--Markovian Effects}
\label{sec31}

We begin with the algebraic case and calculate the squared matrix
gradient of our formula~\eqref{sec1.5}. There are two contributions,
\begin{align}
  \frac{\partial^2}{\partial J^2}{\tr}^2(\Sigma J) &= \frac{\partial^2}{\partial J^2}\tr\Sigma J\tr\Sigma J
                                                   = 2 \frac{\partial}{\partial J}(\tr\Sigma J) \Sigma
                                                   = 2\Sigma^2
 \label{sec3.1}\\
  \frac{\partial^2}{\partial J^2}\tr(\Sigma J)^2 &= \frac{\partial^2}{\partial J^2}\tr\Sigma J\Sigma J
                                                   = 2 \frac{\partial}{\partial J} \Sigma J\Sigma
                                                   = \Sigma^2 + (\tr\Sigma)\Sigma \ ,
 \label{sec3.2}
\end{align}
where the last equality sign is best verified in a tedious, but
straightforward computation in terms of the matrix elements. We thus
find from Eq.~\eqref{eq:AblOpe} for $\nu=2$
\begin{align}
  \left\langle\left(\frac{1}{N} XX^\dagger\right)^2\right\rangle_A &= 
  \frac{M^2}{4N^2\pi^{N(N-1)/4}\prod_{n=1}^N\Gamma(L-(K+n-1)/2)}\nonumber\\
 &\qquad\qquad\qquad   \Big(\Phi_{21}(\Xi)\Sigma^2 + \Phi_{22}(\Xi)\big(\Sigma^2+(\tr\Sigma)\Sigma\big)\Big)
\label{sec3.3}
\end{align}
with the functions
\begin{align}
  \Phi_{21}(\Xi) &= \int_{S > 0} \textrm{d}[S] \exp(-\tr S) {\det}^{{L-(N+K+1)/2}} S \, {\tr}^2(\Theta S^{-1})
  \label{sec3.4a}\\
  \Phi_{22}(\Xi) &= \int_{S > 0} \textrm{d}[S] \exp(-\tr S) {\det}^{{L-(N+K+1)/2}} S \tr(\Theta S^{-1})^2 \ ,
  \label{sec3.4b}
\end{align}
where $\Xi=\Theta$ due to the invariance of the measure. We write out
the traces,
\begin{align}
  {\tr}^2(\Theta S^{-1}) &= \sum_{n=1}^N \Theta_n^2[S^{-1}]_{nn}^2 + 2\sum_{n<m} \Theta_n\Theta_m[S^{-1}]_{nn}[S^{-1}]_{mm}
  \label{sec3.5a}\\
  \tr(\Theta S^{-1})^2 &= \sum_{n=1}^N \Theta_n^2[S^{-1}]_{nn}^2 + 2\sum_{n<m} \Theta_n\Theta_m[S^{-1}]_{nm}^2 \ ,
  \label{sec3.5b}
\end{align}
and observe once more, after reinserting in Eqs.~\eqref{sec3.4a}
and~\eqref{sec3.4b} and pulling out the $\Theta_n$ from the integrals,
that the latter cannot depend on the indices $n$ or $m$ of the matrix
elements $[S^{-1}]_{nm}$ . We just use fixed ones, $n=1$ and $m=2$, do
the remaining sums and arrive at
\begin{align}
  \Phi_{21}(\Xi) &= \Phi_{21}(\Theta) = \frac{\tr\Theta^2}{N}\Psi_d + \frac{{\tr}^2\Theta-\tr\Theta^2}{N(N-1)}\Psi_p
  \label{sec3.6a}\\
  \Phi_{22}(\Xi) &= \Phi_{22}(\Theta) = \frac{\tr\Theta^2}{N}\Psi_d + \frac{{\tr}^2\Theta-\tr\Theta^2}{N(N-1)}\Psi_m
  \label{sec3.6b}
\end{align}
with the integrals
\begin{align}
  \Psi_d &= N\int_{S > 0} \textrm{d}[S] \exp(-\tr S) {\det}^{{L-(N+K+1)/2}} S \, [S^{-1}]_{11}^2
  \label{sec3.7a}\\
  \Psi_p &= N(N-1)\int_{S > 0} \textrm{d}[S] \exp(-\tr S) {\det}^{{L-(N+K+1)/2}} S \, [S^{-1}]_{11}[S^{-1}]_{22}
  \label{sec3.7b}\\
  \Psi_m &= N(N-1)\int_{S > 0} \textrm{d}[S] \exp(-\tr S) {\det}^{{L-(N+K+1)/2}} S \, [S^{-1}]_{12}^2 \ ,
  \label{sec3.7c}
\end{align}
which are independent of $\Xi=\Theta$. As in the case of the first
matrix moment, we managed to decouple the non--Markovian effects from the
matrix integrals to be calculated.

\subsection{Invariant Integrals and Application of the Aomoto Integral}
\label{sec32}

For certain combinations of the unknown integrals, invariant integrals
can be constructed. We simply put $\Xi=\Theta=\mathds{1}_N$ in
Eqs.~\eqref{sec3.6a} and~\eqref{sec3.6b}, use Eqs.~\eqref{sec3.4a}
and~\eqref{sec3.4b} and find
\begin{align}
  \Phi_{21}(\mathds{1}_N) &= \Psi_d + \Psi_p = \int_{S > 0} \textrm{d}[S] \exp(-\tr S) {\det}^{{L-(N+K+1)/2}} S \, {\tr}^2S^{-1}\nonumber\\
                               &\quad = \frac{\pi^{N(N+1)/4}}{N!\prod_{n=1}^N\Gamma(n/2)}
                                             \int_{s > 0} \textrm{d}[s] |\Delta_N(s)| \exp(-\tr s) {\det}^{{L-(N+K+1)/2}} s \, {\tr}^2s^{-1} 
  \label{sec3.8a}\\
  \Phi_{22}(\mathds{1}_N) &= \Psi_d + \Psi_m = \int_{S > 0} \textrm{d}[S] \exp(-\tr S) {\det}^{{L-(N+K+1)/2}} S \tr S^{-2}\nonumber\\
                               &\quad = \frac{\pi^{N(N+1)/4}}{N!\prod_{n=1}^N\Gamma(n/2)}
                                             \int_{s > 0} \textrm{d}[s] |\Delta_N(s)| \exp(-\tr s) {\det}^{{L-(N+K+1)/2}} s \tr s^{-2} \ .
  \label{sec3.8b}
\end{align}
The difference
\begin{align}
  \Psi_p-\Psi_m  &= \frac{\pi^{N(N+1)/4}}{N!\prod_{n=1}^N\Gamma(n/2)}\nonumber\\
                 &\qquad \int_{s > 0} \textrm{d}[s] |\Delta_N(s)| \exp(-\tr s) {\det}^{{L-(N+K+1)/2}} s \left({\tr}^2 s^{-1}-\tr s^{-2}\right) 
  \label{sec3.10}
\end{align}
can be cast into the form of an Aomoto integral which yields
\begin{align}
  \Psi_p-\Psi_m  &= \frac{4N(N-1)\pi^{N(N-1)/4} \prod_{n=1}^N\Gamma(L-(K+n-1)/2)}{(2L-1-(K+N))(2L-(K+N))} \ ,
  \label{sec3.11}
\end{align}
the details are given in~\ref{app1}. The eigenvalue integrals as they
stand in Eqs.~\eqref{sec3.8a} and~\eqref{sec3.8b} can, because of the
squares in the trace terms, not be calculated with the Aomoto
integral, but with other standard techniques, in particular with the
method of integration over alternate variables and the theory of
Pfaffians~\cite{Mehta2004}. Nevertheless the two Eqs.~\eqref{sec3.8a}
and~\eqref{sec3.8b} are not sufficient to obtain all three integrals
$\Psi_d$, $\Psi_p$ and $\Psi_m$ individually. While we managed in
Sec.~\ref{sec2} to reduce the computation of the first matrix moment
to the evaluation of an invariant integral, it is from a more general
viewpoint quite inconceivable that such a reduction is possible to all
orders, otherwise the original problem would be equivalent to an
invariant integral.

\subsection{Extending the Method to Calculate Integrals of Ingham--Siegel Type}
\label{sec33}

We now put forward a method that does not rely on invariant integrals
and allows us to directly compute $\Psi_d$ and $\Psi_p$. Combined with
the result~\eqref{sec3.11}, we obtain all three integrals.  Our method
extends the one given in Refs.~\cite{Siegel_1935,Fyodorov_2002} to
compute the Ingham--Siegel integral~\eqref{eq:IngSieInt}. We begin
with the integral~\eqref{sec3.7a}. Observing that the element of an
inverse matrix can always be expressed as the ratio of the adjugate
and the determinant, we have
\begin{align}
  \Psi_d &= N\int_{S > 0} \textrm{d}[S] \exp(-\tr S) {\det}^{{L-(N+K+1)/2}-2} S \, {\det}^{2} \widetilde{S}_{11} \ ,
  \label{sec3.12}
\end{align}
where the $(N-1)\times (N-1)$ adjugate $\widetilde{S}_{11}$ with
respect to the first row and column appears in the block decomposition
\begin{align}
S&=\begin{bmatrix}
S_{11} &   \widetilde{T}^{\dagger} \\
\widetilde{T}     &   \widetilde{S}_{11} \\
\end{bmatrix}  \qquad \text{with } \qquad \widetilde{T} =
\begin{bmatrix}
S_{12}\\
\vdots \\
S_{1N}
\end{bmatrix}
\end{align}
being an $(N-1)$ component vector.  The determinant of $S$ may then be written as
\begin{align}
  \det S &= \det\widetilde{S}_{11} \left(S_{11}- \widetilde{T}^{\dagger}\widetilde{S}_{11}^{-1} \widetilde{T}\right)
  \label{eq:IngSieDet}
\end{align}
while the trace is given by
\begin{align}
	\tr S &= S_{11} + \tr\widetilde{S}_{11} \ .
  \label{sec3.13}
\end{align}
Since the matrix $S$ is positive definite, the same holds for the
adjugate $\widetilde{S}_{11}$, implying that both determinants are
positive. Hence, the bracket in Eq.~\eqref{eq:IngSieDet} must be
positive as well and we obtain $\widetilde{T}^{\,\dagger}
\widetilde{S}_{11}^{-1} \widetilde{T} < S_{11} < \infty$ as limits for
the $S_{11}$ integration. Collecting everything, we find
\begin{align}
  \Psi_d &= N \inte_{\widetilde{S}_{11}> 0} \text{d}[\widetilde{S}_{11}]\exp(-\tr\widetilde{S}_{11}) {\det}^{{L-(N+K+1)/2}}\widetilde{S}_{11}\nonumber\\
  &\qquad \inte \text{d}[\widetilde{T}] \inte_{\widetilde{T}^\dagger\widetilde{S}_{11}^{-1}\widetilde{T}}^\infty \text{d}S_{11} \exp(-S_{11})
  \left(S_{11} - \widetilde{T}^\dagger\widetilde{S}_{11}^{-1}\widetilde{T}\right)^{L-2-(K+N+1)/2}\nonumber\\
  &=     N \Gamma(L-3/2-(K+N)) \inte_{\widetilde{S}> 0} \text{d}[\widetilde{S}_{11}]\exp(-\widetilde{S}_{11}) {\det}^{{L-(N+K+1)/2}}\widetilde{S}_{11}\nonumber\\
  &\qquad\qquad\qquad\qquad\qquad\qquad \inte \text{d}[\widetilde{T}] \exp(-\widetilde{T}^\dagger\widetilde{S}_{11}^{-1}\widetilde{T})  \nonumber\\
  &=   N\pi^{(N-1)/2} \Gamma(L-3/2-(K+N)) \inte_{\widetilde{S}_{11}> 0} \text{d}[\widetilde{S}_{11}]\exp(-\widetilde{S}_{11}) {\det}^{{L-(N+K)/2}}\tilde{S}_{11}
\label{sec3.14}
\end{align}
The $\tilde{S}_{11}$ integral is again of Ingham--Siegel
type~\eqref{eq:IngSieInt} and
\begin{align}
  \Psi_d &= \frac{4N\pi^{N(N-1)/4}
    \prod_{n=1}^N\Gamma(L-(K+n-1)/2)}{(2L-3-(K+N))(2L-1-(K+N))}
  \label{sec3.14a}
\end{align}
is our final result for $\Psi_d$.

To compute the integral $\Psi_p$, we need the announced extension of the methods
in Refs.~\cite{Siegel_1935,Fyodorov_2002}. We write both matrix elements of the
inverse $S^{-1}$ in terms of their adjugates $\widetilde{S}_{11}$ and $\widetilde{S}_{22}$
which both are $(N-1)\times (N-1)$ matrices,
\begin{align}
  \Psi_p &= N(N-1) \inte_{S > 0} \text{d}[S] \exp(-\tr S) \det S^{L-(K+N+1)/2-2} \det \widetilde{S}_{11}\det\widetilde{S}_{22} \ .
  \label{eq:ZweMomInt3}
\end{align}
To obtain a convenient parameterization of the integration manifold, we introduce
a block decomposition by slicing off the first and the second row and column,
\begin{align}
S=\begin{bmatrix}
\overline{S} & \hat{T}^\dagger \\
\hat{T} & \hat{S} \\
\end{bmatrix}
\qquad \text{with } \qquad \overline{S} =
\begin{bmatrix}
S_{11} & S_{12} \\
S_{12} & S_{22} 
\end{bmatrix}
\label{sec3.15}
\end{align}
where $\overline{S}$ is a $2\times 2$ and $\hat{S}$ an $(N-2)\times (N-2)$ matrix. Moreover,
we set
\begin{align}
\hat{T}_1 &=
\begin{bmatrix}
S_{13} \\
\vdots \\
S_{1N} 
\end{bmatrix} \ , \qquad
\hat{T}_2 =
\begin{bmatrix}
S_{23} \\
\vdots \\
S_{2N} 
\end{bmatrix} \qquad \text{and} \qquad \hat{T} = [\hat{T}_1 \ \hat{T}_2] \ ,
\label{sec3.16}
\end{align}
where $\hat{T}_1$ and $\hat{T}_2$ are $(N-2)$ component vectors and
$\hat{T}$ is an $(N-2)\times 2$ matrix. We also need block
decompositions of the $(N-1)\times (N-1)$ adjugates,
\begin{align}
\widetilde{S}_{11}&=
\begin{bmatrix}
S_{22} & \hat{T}_2^\dagger \\
\hat{T}_2 & \hat{S}
\end{bmatrix} \qquad \text{and} \qquad
\widetilde{S}_{22}=
\begin{bmatrix}
S_{11} & \hat{T}_1^\dagger \\
\hat{T}_1 & \hat{S}
\end{bmatrix} \ ,
\label{sec3.17}
\end{align}
which are consistent with Eqs.~\eqref{sec3.15} and~\eqref{sec3.16}. The block
decompositions~\eqref{sec3.17} lead to the expressions
\begin{align}
  \det \widetilde{S}_{11} &= \det \hat{S} \left( S_{22}- \hat{T}_2^\dagger \hat{S}^{-1} \hat{T}_2\right)
  \label{eq:Det2} \ , \\
  \det \widetilde{S}_{22} &= \det \hat{S} \left( S_{11}- \hat{T}_1^\dagger \hat{S}^{-1} \hat{T}_1\right)
  \label{eq:Det3} \ ,
\end{align}
while the block decomposition~\eqref{sec3.15} yields
\begin{align}
  \det S &= \det \hat{S} \det\left( \overline{S}-\hat{T}^\dagger\hat{S}^{-1} \hat{T}\right) \ ,
  \label{eq:Det1}
\end{align}
which is different as it involves two determinants on the right hand side. The latter is $2\times 2$ and
reads
\begin{align}
\det\left( \overline{S}- \hat{T}^\dagger \hat{S}^{-1} \hat{T}\right)
   &=\left( S_{11}-\hat{T}_1^\dagger \hat{S}^{-1} \hat{T}_1 \right) \left(S_{22}-\hat{T}_2^\dagger \hat{S}^{-1} \hat{T}_2 \right)
  - \left( S_{12}-\hat{T}_1^\dagger \hat{S}^{-1} \hat{T}_2 \right)^2
 \label{eq:Det4} \ .
\end{align}
Since all determinants are positive, we may infer the limits for the integrals over $S_{11}$, $S_{22}$ and $S_{12}$,
\begin{align}
& \hat{T}_1^\dagger \hat{S}^{-1} \hat{T}_1 <  S_{11} <  \infty \qquad \textrm{and} \qquad
	\hat{T}_2^\dagger \hat{S}^{-1} \hat{T}_2 <  S_{22} <  \infty  , \\
& -d +\hat{T}_1^\dagger \hat{S}^{-1} \hat{T}_2 < S_{12} < d +\hat{T}_1^\dagger \hat{S}^{-1} \hat{T}_2 \ ,
\label{sec3.18}
\end{align}
with the short hand notation
\begin{align}
	d &= \sqrt{( S_{11}-\hat{T}_1^\dagger \hat{S}^{-1} \hat{T}_1) (S_{22}-\hat{T}_2^\dagger \hat{S}^{-1} \hat{T}_2)} \ .
\end{align}
Inserting everything in the integral~\eqref{eq:ZweMomInt3}, we obtain
\begin{align}
&\Psi_p = N(N-1) \inte_{\hat{S}> 0} \text{d}[\hat{S}]\exp(-\tr \hat{S}){\det}^{L-(K+N+1)/2}\hat{S} \inte \text{d}[ \hat{T}_1]\inte \text{d}[ \hat{T}_2] \nonumber\\
  &\quad \inte_{\hat{T}_1^\dagger \hat{S}^{-1} \hat{T}_1}^\infty \hspace{-0.4cm} \text{d}S_{11} \exp(-S_{11})
                 \left( S_{11}- \hat{T}_1^\dagger \hat{S}^{-1} \hat{T}_1 \right)
                 \inte_{\hat{T}_2^\dagger \hat{S}^{-1} \hat{T}_2}^\infty \hspace{-0.4cm} \text{d}S_{22} \exp(-S_{22})
                 \left( S_{22}- \hat{T}_2^\dagger \hat{S}^{-1} \hat{T}_2 \right)  \nonumber\\
  &\quad \inte_{-d+\hat{T}_1^\dagger \hat{S}^{-1} \hat{T}_2 }^{d+\hat{T}_1^\dagger \hat{S}^{-1} \hat{T}_2 } \hspace{-0.6cm}\text{d}S_{12}
                 \left( \left( S_{11}-\hat{T}_1^\dagger \hat{S}^{-1} \hat{T}_1 \right) \left(S_{22}-\hat{T}_2^\dagger \hat{S}^{-1} \hat{T}_2 \right)
                 - \left( S_{12}-\hat{T}_1^\dagger \hat{S}^{-1} \hat{T}_2 \right) ^2  \right)^{L-2-\frac{K+N+1}{2}} \nonumber\\
  & \ = N(N-1) \inte_{\hat{S}> 0} \text{d}[\hat{S}]\exp(-\tr \hat{S}) {\det}^{L-(K+N+1)/2} \hat{S}  \nonumber\\
  &\qquad\qquad  \inte \text{d}[ \hat{T}_1]\exp(-\hat{T}_1^\dagger \hat{S}^{-1} \hat{T}_1)\inte \text{d}[ \hat{T}_2] \exp(-\hat{T}_2^\dagger \hat{S}^{-1} \hat{T}_2)\nonumber\\
  &\qquad\qquad \inte_0^\infty \text{d}y_1\exp(-y_1) y_1 \inte_0^\infty \text{d}y_2\exp(-y_2) y_2
                 \inte_{-\sqrt{y_1 y_2}}^{\sqrt{y_1 y_2}} \hspace{-0.4cm} \text{d}x \left( y_1 y_2 - x^2  \right)^{L-2-(K+N+1)/2}
\label{sec3.19}
\end{align}
after obvious shifts of the integration variables $S_{11}$, $S_{22}$
and $S_{12}$. With the rescaling $x \rightarrow \sqrt{y_1y_2}x$, the
integrals over $y_1$, $y_2$ and $x$ become elementary. Moreover, the
$\hat{T}_1$ and $\hat{T}_2$ integrals can be done and we are left with
\begin{align}                
  \Psi_p &= \frac{N(N-1)\pi^{N-3/2}\Gamma^2(L-(K+N)/2)\Gamma(L-1- (K+N+1)/2)}{\Gamma( L-1- (K+N)/2)}  \nonumber\\
         &\qquad\qquad\qquad\qquad \inte_{\hat{S}> 0} \text{d}[\hat{S}]\exp(-\tr\hat{S}) {\det}^{L-(K+N-1)/2} \hat{S} \ ,
\label{sec3.20}
\end{align}
where the remaining integral is, once more, of Ingham--Siegel
type~\eqref{eq:IngSieInt}. Collecting everything we end up with
our final result
\begin{align}    
  \Psi_p &= \frac{4N(N-1) {\pi^{N(N-1)/4}}(2L-2-(K+N))\prod_{n=1}^{N} \Gamma(L-(K+n-1)/2)}{(2L-3-(K+N))(2L-1-(K+N))(2L-(K+N))} 
\label{sec3.21}
\end{align}
for the integral $\Psi_p$.  With Eq.~\eqref{sec3.11}, we also find
\begin{align}    
  \Psi_m &= \frac{4N(N-1) {\pi^{N(N-1)/4}}\prod_{n=1}^{N} \Gamma(L-(K+n-1)/2)}{(2L-3-(K+N))(2L-1-(K+N))(2L-(K+N))} 
\label{sec3.22}
\end{align}
for the integral $\Psi_m$. We notice the relations
\begin{align}    
  \Psi_p &= \frac{(N-1)(2L-2-(K+N))}{2L-(K+N)}\Psi_d
  \qquad \text{and} \qquad
  \Psi_m = \frac{(N-1)}{2L-(K+N)}\Psi_d
\label{sec3.23}
\end{align}
between these integrals.

\subsection{Results}
\label{sec34}

Formulas~\eqref{sec3.6a} and~\eqref{sec3.6b} decouple the matrix
$\Xi$, \textit{i.e.} the non--Markovian effects, from the problem
reducing it to the calculation of the integrals $\Psi_d$, $\Psi_p$ and
$\Psi_m$ which are non--trivial numerical factors.  We insert what we
obtained in the previous Sec.~\ref{sec33} and have
\begin{align}
  \Phi_{21}(\Xi) &= \left(\tr\Xi^2 + \frac{2L-2-(K+N)}{2L-(K+N)}({\tr}^2\Xi-\tr\Xi^2)\right)\frac{\Psi_d}{N}
  \label{sec3.24a}\\
  \Phi_{22}(\Xi) &= \left(\tr\Xi^2 + \frac{1}{2L-(K+N)}({\tr}^2\Xi-\tr\Xi^2)\right)\frac{\Psi_d}{N} 
  \label{sec3.24b}
\end{align}
with $\Psi_d$ given in Eq.~\eqref{sec3.14a}. We plug this into Eq.~\eqref{sec3.3} and arrive at
\begin{align}
  \left\langle\left(\frac{1}{N} XX^\dagger\right)^2\right\rangle_A &= 
  \frac{M^2}{(2L-3-(K+N))(2L-1-(K+N))}\nonumber\\
  &\qquad\qquad  \left(
                       \left(2\frac{\tr\Xi^2}{N^2}+\frac{2L-1-(K+N)}{2L-(K+N)}\, \frac{{\tr}^2\Xi-\tr\Xi^2}{N^2}\right)\Sigma^2 \right.
                                         \nonumber\\
  &\qquad\qquad      + \left. \left(\frac{\tr\Xi^2}{N^2} + \frac{1}{2L-(K+N)}\, \frac{{\tr}^2\Xi-\tr\Xi^2}{N^2}\right)({\tr\Sigma})\Sigma
                                         \right)
\label{sec3.25}
\end{align}
which is our final result for the second matrix moment. Its existence is
guaranteed if the condition
\begin{align}
  L & > \frac{K+N+3}{2}
\label{sec2.26}  
\end{align}
holds, as the derivation shows. The Gaussian case is readily obtained
by applying formulae~\eqref{sec3.1} and~\eqref{sec3.2} to the
generating function~\eqref{sec1.6}, we find
\begin{align}
  \left\langle\left(\frac{1}{N} XX^\dagger\right)^2\right\rangle_G &= 
                  \frac{\tr\Xi^2+{\tr}^2\Xi}{N^2} \, \Sigma^2
                    + \frac{\tr\Xi^2}{N^2} \, ({\tr\Sigma})\Sigma \ .
\label{sec3.27}
\end{align}
In the limit $L,M\rightarrow\infty$ under the condition~\eqref{eq:GauVerR2},
the second matrix moment~\eqref{sec3.25} in the algebraic case
yields the second matrix moment~\eqref{sec3.27} in the Gaussian case,
as it should be.

We finally also provide the results for the matrix variances in the
algebraic and in the Gaussian case. In the former, a straightforward
calculation gives
\begin{align}
  {\textrm{var}_A}\left(\frac{1}{N} XX^\dagger\right) &=
  \frac{M^2(2L+1-(K+N))}{(2L-3-(K+N))(2L-1-(K+N))^2}\nonumber\\
  &\qquad\qquad\qquad  \left(\frac{\tr\Xi^2}{N^2}+\frac{1}{2L-(K+N)}\, \frac{{\tr}^2\Xi}{N^2}\right)\Sigma^2
                                         \nonumber\\
  &\qquad\qquad  +    \frac{M^2}{(2L-3-(K+N))(2L-(K+N))}\nonumber\\
  &\qquad\qquad\qquad  \left(\frac{\tr\Xi^2}{N^2} + \frac{1}{2L-1-(K+N)}\, \frac{{\tr}^2\Xi}{N^2}\right)({\tr\Sigma})\Sigma \ ,
\label{sec3.28}
\end{align}                                       
which coincides with the latter, namely
\begin{align}
  {\textrm{var}_G}\left(\frac{1}{N} XX^\dagger\right) &=
                    \frac{\tr\Xi^2}{N^2} \left( \Sigma^2 + ({\tr\Sigma})\Sigma \right) \ ,
\label{sec3.29}
\end{align} 
in the limit $L,M\rightarrow\infty$ under the condition~\eqref{eq:GauVerR2}.

\section{Conclusions}
\label{sec4}

We studied an algebraic extension of the doubly correlated Wishart
model recently introduced by us, generalizing earlier models by other
authors. In contrast to the Gaussian doubly correlated Wishart model,
it is based on a determinant in the denominator and thus includes in
an expansion all matrix invariants. Our model is motivated by data
featuring algebraic tails which are often not captured by the Gaussian
version, even though mechanisms related to the Central Limit Theorem
work in favor of the latter. The mathematics of our algabraic model is
more demanding than that of the Gaussian version, in particular so as
we consider real random model data matrices, implying real symmetric
correlation and covariance matrices.

We calculated the first and second matrix moments, and thereby
encountered and solved non--trivial mathematical problems, because the
real setup outruled the application of group integrals about which
much less is known in the orthogonal case than in the unitary one. To
circumvent group integrals, we developed an approach that, first,
decouples matrices breaking the rotation invariance from the matrix
integrals, allowing us to reduce them to invariant matrix integrals
which we solve by mapping them onto the Aomoto integral. Nevertheless,
some non--invariant matrix integrals remain which onne can view as
generalizations of the real Ingham--Siegel integral. We succeeded,
second, in calculating them by extending the recursive method for the
standard Ingham--Siegel integral. We hope that our new technical
devlopments might also be helpful in other problems.

Naturally, the question arises if our approach also facilitates the
computation of higher matrix moments. Although we do not see a
fundamental mathematical obstacle, even the computation of the third
matrix moment becomes drastically more involved, which moves tackling
such computations beyond the scope of this contribution. At present,
we do not see competitive alternative methods, but we do definitely
not exclude the possibility that they exist.

\section*{Acknowledgement}

We thank Mario Kieburg and Daniel Waltner for helpful discussions.

\appendix
\label{sec:anhang}

\section{Application of the Aomoto Integral}
\label{app1}

Generalizing the Selberg integral~\cite{Selberg_1944}, the $N$
dimensional Aomoto integral~\cite{Aomoto_1987,Mehta_1997} for positive
constants $a$, $b$ and $\gamma$ as well as $0\leq m \leq N$ is defined as
\begin{align}
  A(a,b,\gamma,N,m) &= \inte_{[0,1]} \produ_{i=1}^N \left(1-u_i \right)^{b-1} u_i ^{a-1} \, \produ_{j=1}^{m}u_j \, |\Delta_N(u)|^{2\gamma} \text{d}[u] \ ,
\label{eq:AomIntFor}
\end{align}
where $[0,1]$ indicates the limits of integration for all variables.
The special choices $m=0$ and $m=N$ are yield two cases of the Selberg
integral with different parameters $a$. The Aomoto integral can be
explicitly calculated,
\begin{align}
A(a,b,\gamma,N,m) &= \produ_{i=0}^{N-1} \frac{\Gamma(a+1+i\gamma)\Gamma(b+i\gamma)\Gamma(1+(i+1)\gamma)}
                                             {\Gamma(a+b+1+(N-1+i)\gamma) \Gamma(1+\gamma)} \nonumber\\
&\qquad\qquad \produ_{j=0}^{N-m-1} \frac{a+b+(N-1+j)\gamma}{a+j\gamma} \ .
\label{eq:AomProFor}
\end{align}
We set $\gamma=1/2$ and change variables according to $u_i=/(1+t_i)$ and find for the integral~\eqref{eq:AomIntFor}
\begin{align}
  A(a,b,1/2,N,m) &=  \inte_{[0,\infty)} \produ_{i=1}^N \frac{t_i ^{a-1}}{(1+t_i)^{a+b+N-1}} \, \produ_{j=1}^{m}\frac{t_j}{1+t_j} \, 
                                                   |\Delta_N(t)| \text{d}[t] \ ,
\label{eq:AomProForErg} .
\end{align}
where we used
\begin{align}
  \left|\Delta\left(\frac{t_1}{1+t_1}, \dots ,\frac{t_N}{1+t_N}\right)\right|&= \produ_{k=1}^N\frac{1}{(1+t_k)^{N-1}} \, |\Delta_N(t)| \ .
\label{aomoto3}  
\end{align}
Employing the further change of variables $t_i=s_i/b$, we consider the limit
\begin{align}
&\lime_{b \rightarrow \infty} b^{N\left(a+1 +\frac{N-1}{2}\right)-(N-m)}A(a,b,1/2,N,m)  \nonumber\\
& \qquad\qquad\qquad  = \lime_{b\rightarrow\infty}\inte_{[0,\infty)} \produ_{i=1}^N \frac{s_i ^{a-1}}{\left(1+\frac{s_i}{b}\right)^{a+b+N-1}} \,
                         \produ_{j=1}^{m}\frac{s_j}{1+\frac{s_j}{b}} \, |\Delta_N(s)| \text{d}[s] \ , \nonumber\\
& \qquad\qquad\qquad  = \inte_{[0,\infty)} \produ_{i=1}^N \text{d}s_i \, \exp(-s_i) s_i ^{a-1} \,
                           \produ_{j=1}^{m}s_j \, |\Delta_N(s)| \text{d}[s] \ .
\label{aomoto4}  
\end{align}
On the other hand, we obtain in the same limit from the result~\eqref{eq:AomProFor}
\begin{align}
&\lime_{b \rightarrow \infty} b^{N\left(a+1 +\frac{N-1}{2}\right)-(N-m)}A(a,b,1/2,N,m)  \nonumber\\
& \qquad\qquad\qquad = \produ_{i=0}^{N-1} \frac{\Gamma(a+1+i/2)\Gamma((3+i)/2)}{\Gamma(3/2)} \, \produ_{j=0}^{N-m-1} \frac{1}{a+j/2} \ .
 \label{aomoto5}  
\end{align}
Combining, Eqs.~\eqref{aomoto4} and~\eqref{aomoto5} we find the integration formula
\begin{align}
&  \inte_{[0,\infty)} \produ_{i=1}^N  \exp(-s_i) s_i ^{a-1} \, \produ_{j=1}^{m}s_j \, |\Delta_N(s)| \text{d}[s]  \nonumber\\
&\qquad\qquad\qquad  =  \produ_{i=0}^{N-1} \frac{\Gamma(a+1+i/2)\Gamma((3+i)/2)}{\Gamma(3/2)} \, \produ_{j=0}^{N-m-1} \frac{1}{a+j/2} \ .
 \label{aomoto6}
\end{align}
For the first matrix moment, we need the case $m=N-1$, 
\begin{align}
  \inte_{[0,\infty)} \produ_{i=1}^N  \exp(-s_i) s_i ^{a-1} \, \produ_{j=1}^{N-1}s_j \, |\Delta_N(s)| \text{d}[s]
     &=  \frac{1}{a} \produ_{i=0}^{N-1} \frac{\Gamma(a+1+i/2)\Gamma((3+i)/2)}{\Gamma(3/2)} \ ,
 \label{aomoto7}
\end{align}
because writing $\tr s^{-1}$ out as a sum yields for the integral in Eq.~\eqref{sec2.8}
\begin{align}
& \frac{1}{N} \sum_{n=1}^N \int_{s > 0} \textrm{d}[s] |\Delta_N(s)| \exp(-\tr s) {\det}^{{L-(N+K+1)/2}} s \, \frac{1}{s_n} \nonumber\\
  &\qquad\qquad\qquad = \int_{s > 0} \textrm{d}[s] |\Delta_N(s)| \exp(-\tr s) {\det}^{{L-(N+K+1)/2}} s \, \frac{1}{s_N} \nonumber\\
  &\qquad\qquad\qquad = \int_{s > 0} \textrm{d}[s] |\Delta_N(s)| \exp(-\tr s) \produ_{n=1}^N s_n^{{L-(N+K+1)/2-1}} \,  \produ_{n=1}^{N-1}s_n \ .
 \label{aomoto7a}
\end{align}
Thus, $a=L-(N+K+1)/2$. For the second matrix moment, we have to put
$m=N-2$,
\begin{align}
  \inte_{[0,\infty)} \produ_{i=1}^N  \exp(-s_i) s_i ^{a-1} \, \produ_{j=1}^{N-2}s_j \, |\Delta_N(s)| \text{d}[s]
     &=  \frac{1}{a(a+1/2)} \produ_{i=0}^{N-1} \frac{\Gamma(a+1+i/2)\Gamma((3+i)/2)}{\Gamma(3/2)} \ ,
 \label{aomoto8}
\end{align}
since the integral in Eq.~\eqref{sec3.10} may be written as
\begin{align}
& 2 \sum_{n<m} \int_{s > 0} \textrm{d}[s] |\Delta_N(s)| \exp(-\tr s) {\det}^{{L-(N+K+1)/2}} s \, \frac{1}{s_ns_m}  \nonumber\\
& \qquad\qquad\qquad = N(N-1)\int_{s > 0} \textrm{d}[s] |\Delta_N(s)| \exp(-\tr s) {\det}^{{L-(N+K+1)/2}} s \, \frac{1}{s_Ns_{N-1}}  \nonumber\\
& \qquad\qquad\qquad = N(N-1)\int_{s > 0} \textrm{d}[s] |\Delta_N(s)| \exp(-\tr s) {\det}^{{L-(N+K+1)/2-1}} s \, \produ_{n=1}^{N-2}s_n \ ,
 \label{aomoto8a}
\end{align}
and we again have to set $a=L-(N+K+1)/2$.  Formula~\eqref{aomoto6} may
also be obtained differently, but the present derivation seems to be a
very convenient one.

%
%
%
%


\end{document}